\documentclass[12pt,preprint]{aastex}

\newcommand{\nab}{\mathbf{\nabla}}
\newcommand{\magf}{\mathbf{B}}
\newcommand{\magfp}{\mathbf{b}}
\newcommand{\cur}{\mathbf{J}}
\newcommand{\vel}{\mathbf{u}}
\newcommand{\pot}{\mathbf{A}}

\shorttitle{Shear-driven instabilities in Hall-MHD plasmas}
\shortauthors{Bejarano, G\'omez \& Brandenburg}

\begin{document}


\title{Shear-driven instabilities in Hall-MHD plasmas}


\author{Cecilia Bejarano and Daniel O. G\'omez\altaffilmark{1}}
\affil{Instituto de  Astronom\'ia y F\'isica del Espacio\\ 
(Consejo Nacional de Investigaciones Cient\'ificas y T\'ecnicas - Universidad de Buenos Aires),\\
Ciudad Universitaria, 1428, C.A.B.A, Buenos Aires, Argentina}
\email{cbejarano@iafe.uba.ar - gomez@iafe.uba.ar}

\and

\author{Axel Brandenburg\altaffilmark{2}}
\affil{NORDITA, AlbaNova University Center,\\
Roslagstullsbacken 23, SE-10691 Stockholm, Sweden}
\email{brandenb@nordita.org}


\altaffiltext{1}{Also at Departamento de F\'isica, Facultad de Ciencias Exactas y Naturales, Universidad de Buenos Aires, Ciudad Universitaria, 1428, C.A.B.A, Buenos Aires, Argentina}
\altaffiltext{2}{Also at Department of Astronomy, Stockholm University, SE-10691 Stockholm, Sweden}


\begin{abstract}

The large-scale dynamics of plasmas is well described within the framework of magnetohydrodynamics (MHD). However, whenever the ion density of the plasma becomes sufficiently low, the Hall effect is likely to become important. The role of the Hall effect has been studied in several astrophysical plasma processes, such as magnetic reconnection, magnetic dynamo, MHD turbulence or MHD instabilities. In particular, the development of small-scale instabilities is essential to understand the transport properties in a number of astrophysical plasmas. The magneto-rotational instability, which takes place in differentially rotating accretion disks embedded in relatively weak magnetic fields, is just one example. The influence of the large-scale velocity flows on small-scale instabilities is often approximated by a linear shear flow. In this paper we quantitatively study the role of the Hall effect on plasmas embedded in large-scale shear flows. More precisely, we show that an instability develops when the Hall effect is present, which we 
therefore term as the {\it Hall magneto-shear} instability. As a particular case, we recover the so-called 
magneto-rotational instability and quantitatively assess the role of the Hall effect on its development and 
evolution. 

\end{abstract}


\section{Introduction}\label{sec:intro}
The large-scale dynamics of astrophysical plasmas is theoretically described within the framework of magnetohydrodynamics (MHD). In many cases of interest, these astrophysical flows are characterized by extremely large Reynolds numbers, which in turn implies that a wide range of spatial scales are relevant to properly describe their dynamical behavior. At sufficiently small spatial scales, kinetic plasma processes might become non-negligible under certain circumstances. For instance, in a fully ionized plasma, whenever one reaches spatial scales as small as the ion skin depth $c/\omega_{pi}$ (where $c$ is the speed of light and $\omega_{pi}$ is the ion plasma frequency), the Hall effect should not be 
neglected. This regime correspond to fully ionized plasmas with sufficiently low ion densities. However, it can also arise in cold plasmas with a low ionization fraction $\chi$ in wich case the relevant Hall scale is given by $(c/ \omega_{pi})\chi ^{-1/2}$ (see \citet{pandey} for details on Hall-MHD of partially ionized plasmas). In cold plasmas such as those present in protoplanetary disks, another kinetic effect known as ambipolar diffusion might become relevant \citep{brazwe}, especially toward the disk surface \citep{pandey}. The relative importance of non-ideal effects such as Hall, ohmic dissipation and ambipolar diffusion has been extensively discussed by \citet{balter} as well as by \citet{pandey}. To highlight the relevance of the Hall effect in astrophysical plasmas, it is useful to compare the orders of magnitude of the ohmic ($O$), inductive ($I$), and Hall ($H$) terms in the generalized Ohms's law. For example, in a typical protostellar disk, the relevant ratios are  $H/O \sim 10^{2}$ and $H/I \sim 10^{4}$ \citep{balter}; while, for dwarf nova disk $H/I \sim 1$ \citep{sanostone1} and for the crust of a neutron star $H/O \sim 10^{3}$ \citep{holrud}.

There are many examples of astrophysical plasma flows for which the role of the Hall effect has been studied: regions of star formation \citep{norhey}, dense molecular clouds  \citep{warng}, the interstellar medium \citep{spa, kin} or even the early universe \citep{taj}. Due to their intense magnetic fields, the Hall currents can be relevant also in white dwarfs and neutron stars \citep{urpyak, shaurp, pot}. Also, the role of Hall currents in the generation of magnetic fields by turbulent dynamo activity has been studied by \citet{min1}; see also  \citet{min2,min3} and references therein.

Even though the dynamics of small-scale structures is often unobservable in astrophysical flows, they may 
play an important role through nonlinear interactions with the large-scale part of the flow. In many cases, the small-scale dynamics of the fluids are instrumental in changing the transport properties of the large-scale dynamics of the fluids, and 
therefore it is relevant to identify potential instabilities in the microscale. At these small spatial scales, the large scale velocity field can be reasonably approximated by a linear shear flow. The so-called shear-driven 
instabilities are those that originate as a result of the presence of a large-scale velocity shear.

In this paper we study the potential relevance of the Hall effect in the presence of an external magnetic field as well as a linear shear flow. In particular, we focus our attention on the following two types of flows with Hall effect: (a) non-rotating shear flows leading to what we call Hall magneto-shear instability (Hall-MSI); and (b)   
differententially rotating flows leading to Hall magneto-rotational instability (or Hall-MRI). In the absence of both rotation and shear, the linear modes in Hall-MHD correspond to right-hand polarized {\it whistlers}  and left-hand polarized {\it ion-cyclotron} waves (see for instance \citet{mahajan05} and references therein). When these modes propagate embedded in a shear flow, the ion-cyclotron mode might become unstable. This instability takes place when the shear is steep enough to be larger than the ion-cyclotron frequency. A linear analysis of this instability has been recently reported by \citet{kunz}, for the case of weakly ionized plasmas. Another study on the influence of the Hall effect on weakly ionized plasmas subjected to differential rotation, was reported by \citet{rudkit} for finite magnetic Reynolds numbers up to $300$ (see also \citet{rudsha}). 

In the present study we adopt a one-dimensional configuration, perform a linear analysis to identify potential instabilities and then compare with numerical simulations. The set of equations as well as the simplifying assumptions that we adopt are listed in \S\ref{sec:eqs}. The dispersion relation for the linear regime is shown in \S\ref{sec:disprel}. We briefly describe the numerical code employed in \S\ref{sec:simul}. The role of the Hall effect on non-rotating shear flows is presented in \S\ref{sec:hmsi}, while the action of Hall
currents on the well known magneto-rotational instability is discussed in \S\ref{sec:hmri}. The non-linear behavior is tackled through a qualitative approach in \S\ref{sec:nonlinear}. Finally, we summarize our conclusions in \S\ref{sec:conclu}. 
\section{General Equations}\label{sec:eqs}
The dimensionless Hall-MHD equations in a rotating reference frame with angular velocity 
$\mathbf{\Omega}=\Omega_0 \mathbf{\hat {z}}$ are listed below. We use $t_{0}$ as our time unit, velocities 
are in units of the Alfv\'en velocity $v_A$, and particle densities are in units of $n_0$. For a fully 
ionized hydrogen plasma, the continuity equation is
\begin{equation}
\frac{\partial n}{\partial t} + \nab \cdot (n \vel)=0\ ,
\label{eq:cont}
\end{equation}
where $n$ is both the proton and the electron particle density (i.e.\ $n_e = n_i = n$) to guarantee 
charge neutrality. The equation of motion for this plasma is
\begin{equation}
  \frac{\partial \vel}{\partial t}+(\vel \cdot \nab)\vel=-2 \beta \nab h +\frac{\nab \times \magf \times
\magf}{n}-2\Omega\mathbf{\hat{z}} \times \vel + \nu\nab^2\vel \ ,
\label{eq:ns}
\end{equation}
where $\beta=(c_{s}/v_{A})^{2}$ is the plasma parameter (being $c_{s}$ the sound speed), $h$ is the enthalpy density for a barotropic flow for each species (electrons or protons), $\Omega = \Omega_0 t_0$, and $\nu$ is the dimensionless viscosity 
coefficient. The induction equation is
\begin{equation}
\frac{\partial \pot}{\partial t}=  \varepsilon \beta \nab h - \nab \Phi  
+ \left( \vel -\varepsilon\frac{\nab \times \magf}{n} \right)\times \magf -\eta\nab\times\magf \ ,
\label{eq:ind}
\end{equation}
where $\pot$ is the vector potential, $\Phi$ is the electrostatic potential, 
$\varepsilon=c /( \omega_{pi}l_0)$ is the Hall parameter (being $l_0=v_A t_{0}$ our length unit), and $\eta$ is the dimensionless electric resistivity. The expresion for the enthalpy density for an ideal and isothermal gas (i.e.\ $T_{e}=T_{i}=T=const.$) is
\begin{equation}
h_{e,i}= h = \ln n\ .
\label{eq:enthalpy}
\end{equation}

In the co-rotating reference frame, centered at $r=r_{0}$, we assume a local cartesian small box (with side size $\Delta \ll r_{0}$) such that $x$ is oriented in 
the radial 
direction and $y$ is along the azimuthal direction (i.e.\ $\mathbf{\hat{r}} \rightarrow \mathbf{\hat{x}}$ and $\mathbf{\hat{\phi}} 
\rightarrow \mathbf{\hat{y}}$).

Assuming an external magnetic field $B_{0}\mathbf{\hat{z}}$ and an externally applied linear shear flow to a small parcel of plasma
located at a radial distance $r_0$, the velocity and the magnetic vector
fields can be written in terms of the three-dimensional small-scale fields $\vel$ and $\magfp$ as follows: 
\begin{equation}
\quad 
\magf  =B_{0}\mathbf{\hat{z}}+ \magfp(z,t)
\ ,\quad 
\vel  = - s x\mathbf{\hat{y}}+\vel(z,t)
\ ,
\label{eq:1d}
\end{equation}
where 
\begin{equation}
s=-t_0\partial_x u_y
\label{eq:shear}
\end{equation}
is the externally applied and constant shear. The 
external magnetic field $ B_0\mathbf{\hat{z}}$ ($B_{0}=1$ in the dimensionless version) is distorted by the applied shear flow, developing 
perpendicular components on both the velocity and the magnetic vector fields.

We adopt the shearing-box approximation (see \citet{haw} and \citet{umureg} for details on the shearing-box approach). The dimensionless Hall-MHD shearing-box equations are the following:
\begin{equation}
(\partial_{t}-sx\partial_{y})n+\nab \cdot (n \vel)=0 \ ,
\label{ec:sbox1}
\end{equation}
\begin{equation}
(\partial_{t}-sx\partial_{y})\vel + (\vel \cdot \nab)\vel=-2 \beta \nab h + \frac{\nab \times \magf \times \magf}{n}-2\Omega \mathbf{\hat{z}} \times \vel + su_{x}\mathbf{\hat{y}} +\nu\nab^2\vel\ ,
\label{ec:sbox2}
\end{equation}
\begin{equation}
(\partial_{t}-sx\partial_{y})\magf=\nab \times \left( \vel -\varepsilon\frac{\nab \times \magf}{n} \right)\times \magf -sB_{x}\mathbf{\hat{y}}+\eta\nab^2\magf \ .
\label{ec:sbox3}
\end{equation}

From this set of equations, we analyze the influence of the Hall term in two cases of interest: in \S\ref{sec:hmsi} we study non-rotating ($\Omega=0$ and $s\neq 0$) plasmas embedded in a large scale shear flow; in \S\ref{sec:hmri} we study differentially rotating disks ($\Omega \neq 0$ and $s=a\Omega$ where $a$ comes from a generic profile of differential rotation given by $\Omega (r) = \Omega_0 (r/r_0)^{-a}$). 
\section{Dispersion Relation}\label{sec:disprel}
After linearising the Hall-MHD equations with the one-dimensional geometric setup described in the previous section by Eq.~(\ref{eq:1d}), the system splits into two subsystems. One of them corresponds to the longitudinal modes associated to sound waves propagating 
along the magnetic field, i.e.\
\begin{equation}
\partial_{t} \delta n  = -\partial_{z} u_{z} \ ,
\label{eq:lincont}
\end{equation}
\begin{equation}
\partial_{t} u_{z}  = -2 \beta \partial_{z} \delta n \ , 
\label{eq:linnsz}
\end{equation}
where $\delta n$ are small perturbations from a spatially uniform particle density, i.e.\ $n = 1 + \delta n$. 
The other subsystem corresponds to the perpendicular degrees of freedom described by
\begin{equation}
\partial_{t}\vel_{\perp}  = \partial_{z}\magfp_{\perp} + 
\left(
\begin{array}{cc}
0 & 2\Omega \\
(s-2\Omega) & 0
\end{array} 
 \right) 
\vel_{\perp} \ ,
\label{eq:linnsperp}
\end{equation}
\begin{equation}
\partial_{t}\magfp_{\perp}  =\partial_{z}\vel_{\perp}+
\left(
\begin{array}{cc}
0 & 0 \\
-s & 0
\end{array} 
 \right) 
\magfp_{\perp}
+\varepsilon 
\left(
\begin{array}{cc}
0 & 1 \\
-1 & 0
\end{array}
\right) 
\partial_{zz}^{2}\magfp_{\perp}\ .
\label{eq:linind}
\end{equation}
Equations (\ref{eq:lincont})-(\ref{eq:linind}) show that the perpendicular part $(\vel_{\perp}, \magfp_{\perp})$ of the linear dynamics 
(i.e.\ Eqs.~(\ref{eq:linnsperp})--(\ref{eq:linind})) remain fully decoupled from the 
longitudinal  part $({\delta n, u_z})$ (i.e.\ Eqs.~(\ref{eq:lincont})--(\ref{eq:linnsz})), which is responsible for the 
propagation of acoustic waves.

Assuming that the components of the perpendicular modes $\vel_{\perp}$ and $\magfp_{\perp}$ are proportional to  $e^{i(kz-\omega t)}$, this set of equations leads to the 
following dispersion relation:
\begin{equation}
\omega^{4}-2C_{2}\omega^{2}+C_{0}=0\ ,
\label{eq:reldisp}
\end{equation}
where
\begin{equation}
C_{2}(k) =\frac{\varepsilon ^{2}}{2}k^{4}+\left(1-\frac{\varepsilon s}{2}
\right)k^{2} + \Omega(2\Omega-s) \ ,
\label{eq:c2}
\end{equation}
\begin{equation}
C_0(k) = k^2\left[1+\varepsilon(2\Omega-s)\right]\left[k^2(1+2\varepsilon
\Omega)-2s\Omega\right]\ .
\label{eq:c0}
\end{equation}
Note that this particular dispersion relation is a bi-quadratic polynomial with coefficients depending 
on $k^2$. Therefore, the solutions of the dispersion relation can be written as
\begin{equation}
\omega _\pm^{2} = C_2\ \pm\ \sqrt{C_2^2\ -\ C_0} \ .
\label{eq:omegapm}
\end{equation}

The set of relevant parameters arising in the linearized Hall-MHD equations (Eqs.~(\ref{ec:sbox1})--(\ref{ec:sbox3})) are: the mode wavenumbers (the problem depends on $k^{2}$), the Hall parameter $\varepsilon$, the rotation frequency $\Omega$, the local shear $s$ which is the $z$-component of the vorticity (i.e.\ $\mathbf{w}=\mathbf{\nab}\times \vel=(0,0,s)$), and the external magnetic field in units of the Alfv\'enic velocity $v_{A}$. The rotation frequency, the local vorticity, and the external magnetic field are actually vector quantities ($\mathbf{\Omega}$, $\mathbf{w}$, and $\mathbf{B}$), which in this 1D version
means that they can only be aligned or anti-aligned to one another. If we choose
the magnetic field along $\mathbf{\hat{z}}$, there are four possible cases, depending on the orientations of the local vorticity $\mathbf{w}$ and the rotation frequency $\mathbf{\Omega}$. Up until now, the time unit $t_{0}$ remained undetermined. Therefore, we choose $t_{0}$ so that $|\partial_{x}u_{y}|=1$, and therefore $s=\pm1$. We explore the branches $\omega_{+}^{2}$ and $\omega_{-}^{2}$ in the space of parameters given by $\varepsilon$, $\Omega$, and $k^{2}$. We find that only $\omega_{-}^{2}$ can lead to an instability whenever $\omega_{-}^{2}<0$. 
\begin{figure}
\epsscale{.8}
\plotone{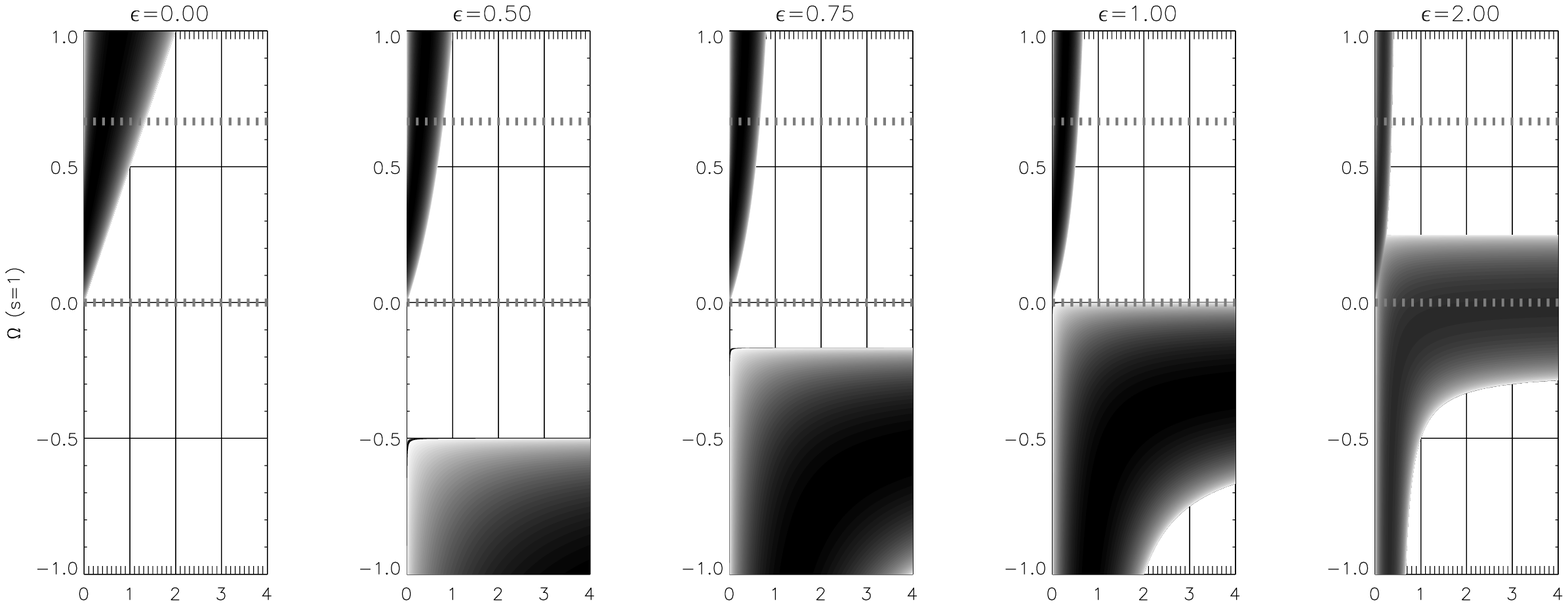}
\plotone{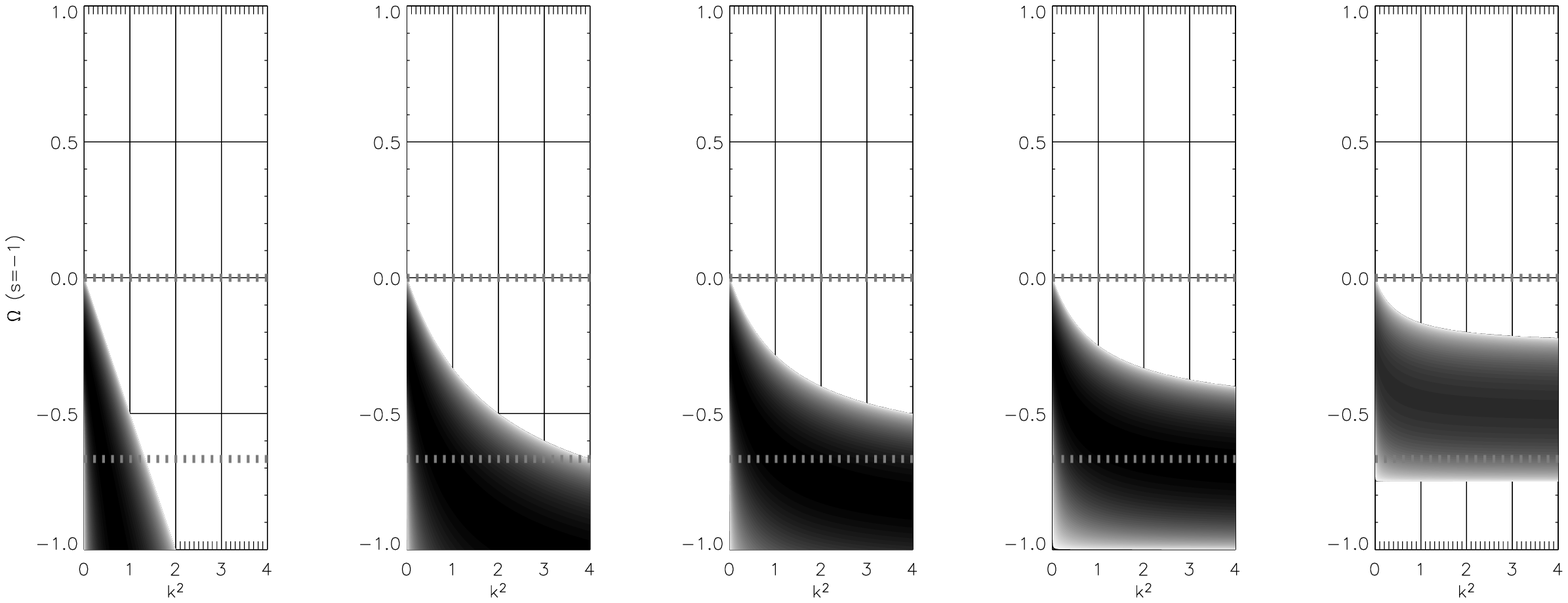}
\caption{Contour levels of the growth rate for different Hall parameter values on the
($\Omega, k^2$) plane for positive (s = +1, top panel) and negative shear (s = -1, bottom panel). Stable regions are white and darker shades correspond to larger instability growth rates. The keplerian ($\Omega=s/a$) and the non-rotational ($\Omega=0$) cases are shown by the horizontal dotted gray lines. } 
\label{contours}
\end{figure}
%

%
In Fig.~\ref{contours} we show a sequence of plots of the growth rate $\gamma_{-}$ (with $\gamma^{2}_{-}=-\omega_{-}^{2}$) in the parameter plane $(\Omega, k^{2})$ for several values of
the Hall parameter. The upper panels of Fig.~\ref{contours} correspond to $s=1$ and  the lower panels correspond to $s= -1$.  Note that $\Omega$ can be either positive or negative. The horizontal dotted gray lines represent the two cases of interest which we analyze in the following sections: non-rotating flows ($\Omega=0$ and $s \ne 0$) and differentially rotating keplerian flows ($\Omega \ne 0$ and $s=a\Omega$ where a=3/2;  in this set of units $\Omega=s/a=\pm 2/3$). As shown in Fig.~\ref{contours}, if the shear is aligned with the magnetic field (positive shear), the non-rotating case becomes unstable for $\varepsilon > 1$. In particular, all wavenumbers are unstable for $s=1$ and $\varepsilon > 0$ showing an asymptotic behavior for $k^{2} \rightarrow \infty$. On the other hand, if the shear is anti-aligned ($s=-1$) there are no unstable modes for the non-rotating case. Meanwhile the keplerian case always has unstable wavenumbers, but they are modified by the different values of the Hall parameter.  Figure \ref{contours} clearly shows that the $s>0$ and $s<0$ cases are entirely different, except for the purely HMD case ($\varepsilon=0$). In all cases considered, there is an interval of rotation frequencies for which all wavenumbers are unstable. In the cases with positive shear ($s=1$), the unstable strip is given by $-1/2 \varepsilon < \Omega < (\varepsilon-1)/2 \varepsilon$. For the cases with negative shear ($s=-1$), is given by $-(\varepsilon-1)/2 \varepsilon \leq \Omega \leq-1/2 \varepsilon$.
\section{Numerical simulations}\label{sec:simul}
A natural first step is to test the growth rates arising from the dispersion
relation with the numerical results coming out from simulations. We
have explored two different cases: non-rotating shear flows in \S\ref{sec:hmsi} and differentially rotating flows (which can locally be approximated by a linear shear flow) in \S\ref{sec:hmri}. We performed hydromagnetic compressible simulations in one dimension using the {\sc Pencil Code}\footnote{http://code.google.com/p/pencil-code} \citep{bradob}, 
a high-order finite-difference numerical suite for compressible magnetohydrodynamic flows.

The simulations have been done with externally imposed shear, either with or without rotation. For the 
initial condition, we assume a monochromatic Alfv\'en wave travelling in $\mathbf{\hat{z}}$ with a
very small amplitude. The resistivity and viscosity coefficients (i.e.\ $\eta$ and $\nu$)  were taken small 
enough to be 
non-negligible only at the smallest spatial scales. The computational domain is $2 \pi$-periodic 
in the $\mathbf{\hat{z}}$ direction, which is both the rotation axis and the orientation of the external magnetic field.
\section{Hall Magneto-Shear Instability (Hall-MSI)}\label{sec:hmsi}
From equations (\ref{eq:c2})-(\ref{eq:c0}) we derive the corresponding coefficients for the case of non-rotating plasmas (i.e.\ $\Omega_{0}=0$) embedded in large-scale shear flows , i.e.\
\begin{equation}
C_2  =\frac{\varepsilon^2 k^4}{2} + \left( 1-\frac{\varepsilon}{2} \right) k^2 \ , 
\label{c2_hmsi}
\end{equation}
\begin{equation}
C_{0}  =k^4 (1-\varepsilon) \ .
\label{c0_hmsi}
\end{equation} 

Note that Eqs.~(\ref{eq:c2})--(\ref{eq:c0}), for the particular case $\Omega =0$, depend on the combination $\varepsilon s$, with $s=\pm 1$ (depending on the relative orientation of the shear's vorticity with respect to the external magnetic field). In Eqs.~(\ref{c2_hmsi})--(\ref{c0_hmsi}) we simply replaced $\varepsilon s$ as a ``signed $\varepsilon$''. Positive values of $\varepsilon$ correspond to a shear-related vorticity $\mathbf{w}$ aligned with the external magnetic field, while negative values of $\varepsilon$ describe anti-aligned configurations.

These coefficients are now considerably simpler and allow a straightforward study of this instability.
Note that the even simpler case of $\varepsilon = 0$ is stable, since $C_2 = k^2$ and 
$C_0 = k^4$, which therefore describes the propagation of Alfv\'en waves along the $\mathbf{\hat{z}}$ 
direction. Therefore, when the Hall current is neglected, an external shear flow is unable to drive 
an instability, as expected. On the other hand, it is straightforward to prove that $C_2^2 - C_0 \geq 0$.  
Consequently, $\omega_{\pm}^{2}$ are real numbers (see Eq.~(\ref{eq:omegapm})) and there are a priori four possible cases depending on the signs of $C_{0}$ and $C_{2}$. However, there is no region on the $(k^{2}, \varepsilon)$ plane where $C_{0}>0$ and $C_{2}<0$ simultaneously. Therefore the condition for instability reduces to $C_0 < 0$, which in turn 
implies $\varepsilon > 1$. Considering the units used for the present study, $\varepsilon > 1$ 
corresponds to $|\partial_x u_y| > \omega_{ci}$ (where $\omega_{ci} = eB_0/(m_i c)$ is the ion-cyclotron 
frequency), which corresponds to a rather steep linear shear flow, except perhaps for considerably 
mild external magnetic fields.
\begin{figure}
\epsscale{.4}
\plotone{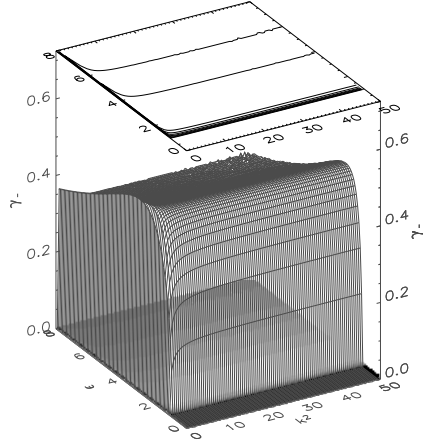}
\caption{Growth rate of the unstable branch as a function of wavenumber and Hall parameter
 (i.e.\ $\gamma_{-}(k^2, \varepsilon)$) for a shear flow ($s=1$)
without rotation. A contour plot is overlaid at the top and a gray scale image
is shown at the bottom.} 
\label{growth_hmsi}
\end{figure}
%

%
In Fig.~\ref{growth_hmsi} we show the growth rate for the unstable branch $\gamma_- 
(k^2, \varepsilon)$. The instability region extends all the way from $\varepsilon = 1$ to 
$\varepsilon \rightarrow \infty$ and, more importantly, it is unstable for all wavenumbers. At 
large wavenumbers, the growth rate reaches its maximum value $\gamma_{-}^{\max} = 0.5$ at $\varepsilon = 2$
(i.e.\ $|\partial_x u_y| > 2\omega_{ci}$). The contour levels of the growth rate $\gamma_-$ on the $(k^2, \varepsilon)$ diagram, which are overlaid at the top, clearly show an asymptotic behavior toward $k\rightarrow\infty$. This implies that all Fourier modes are unstable for a specific range of values of the Hall parameter.
\begin{figure}
\epsscale{.4}
\plotone{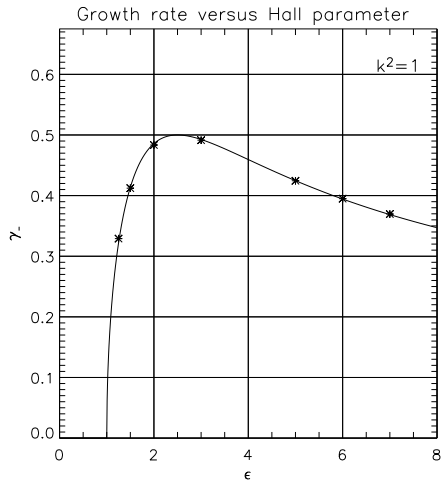}
\caption{Overlap between the theoretical model (solid line) and the
simulations ($\ast$) for $k=1$ for a shear flow ($s=1$)
without rotation.}
\label{overlap_hmsi}
\end{figure}
\begin{figure}
\epsscale{.8}
\plotone{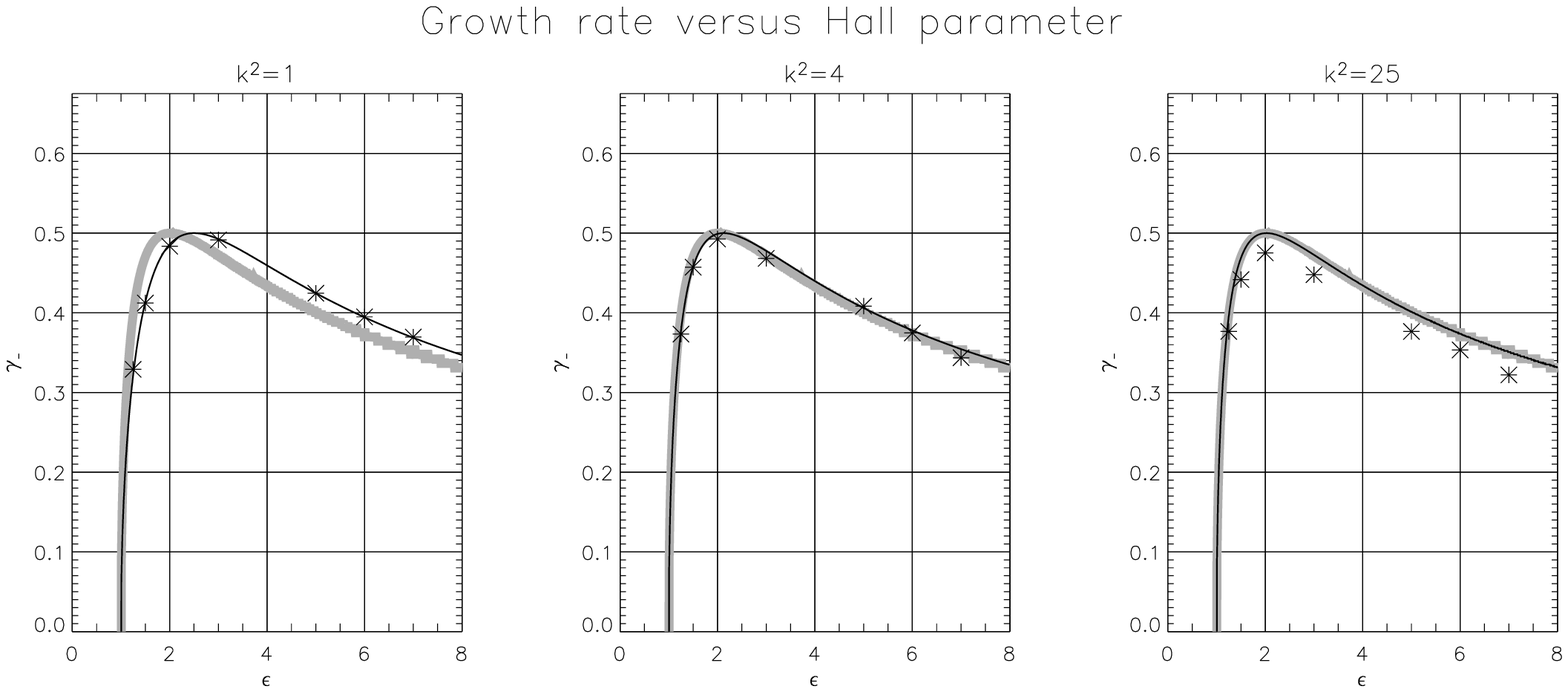}
\caption{Overlap between the theoretical model (solid line) and the
simulations ($\ast$) for $k=1$, $k=2$, and $k=5$.  The asymptotic behavior for
$k^{2} \rightarrow \infty$ (gray line) is indicated  in the three cases.} 
\label{asym_hmsi}
\end{figure}
%
In Fig.~\ref{overlap_hmsi} 
a transverse cut of $\gamma_{-}$ vs. $\varepsilon$ is shown for $k^2 = 1$. There is also a very good 
agreement between the analytical result (solid line) and numerical results ($\ast$) obtained from 
simulations. The instability region corresponds to positive values of $\varepsilon$ which implies that $\mathbf{s}$ and $\magf$ are aligned.
In Fig.~\ref{asym_hmsi} we show profiles of the growth rate as a function of the Hall parameter for three different wavenumbers ($k=1$, $k=2$, and $k=5$) in order to display the asymptotic behavior when $k^{2} \rightarrow \infty$, represented by a gray curve in each panel. Note that in the asymptotic limit, the frequency of the unstable mode becomes gradually independent of $k$. This causes the convective and Hall terms on Eq.~(\ref{eq:linind}) (i.e. the first and third terms on the right hand side) grow like $k^2$, while the remaining terms become comparatively negligible. Therefore, the large-$k$ asymptotic regime is in fact approximately independent of $k$.

\begin{figure}
\epsscale{.4}
\plotone{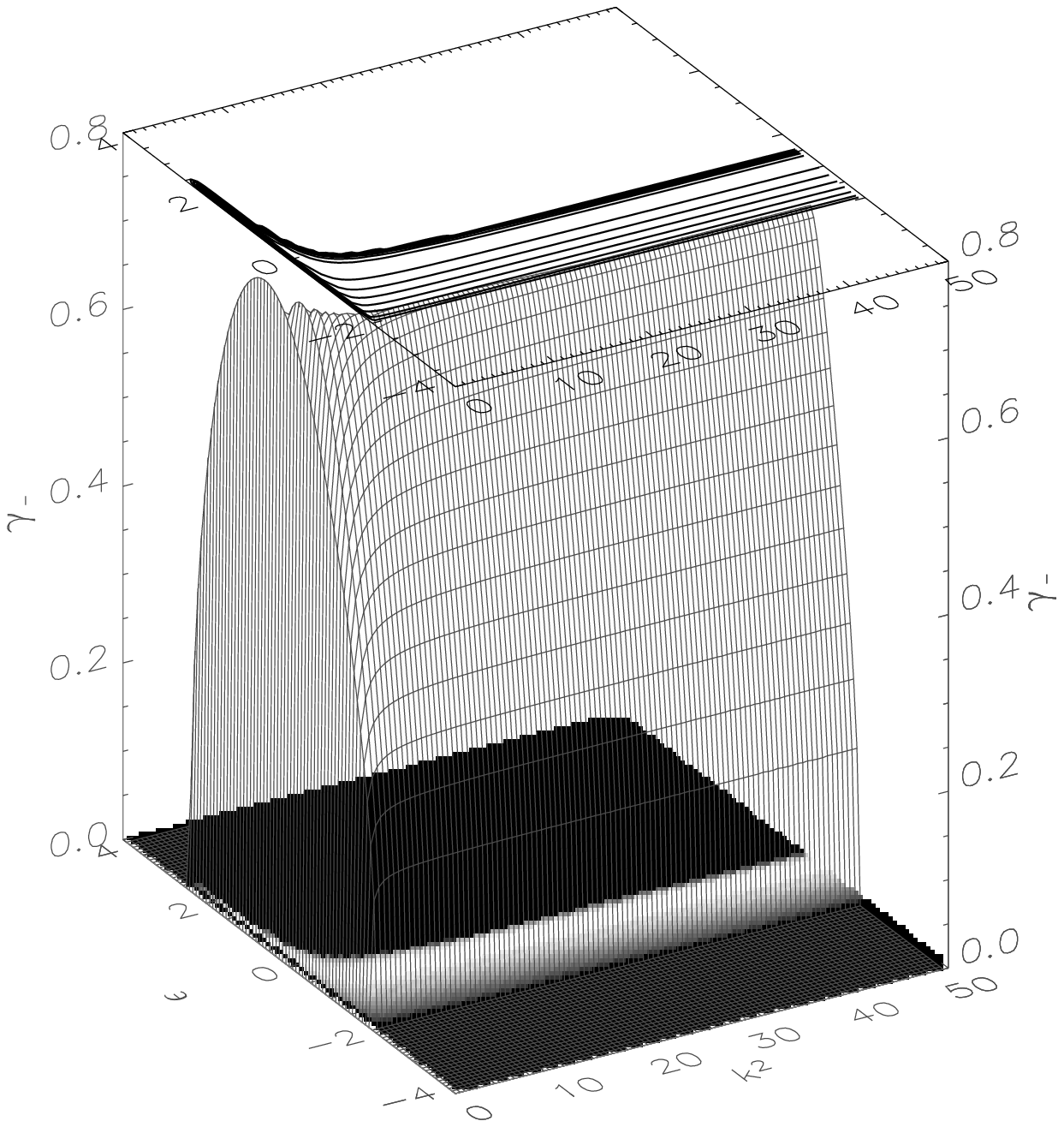}
\caption{Growth rate of the unstable branch as a function of wavenumber and Hall parameter (i.e.\ $\gamma_{-}(k^2,\varepsilon)$) for a keplerian rotation
profile. A contour plot is overlaid at the top and a gray scale image is shown at the
bottom.} 
\label{growth_hmri}
\end{figure}
\begin{figure}
\epsscale{.8}
\plotone{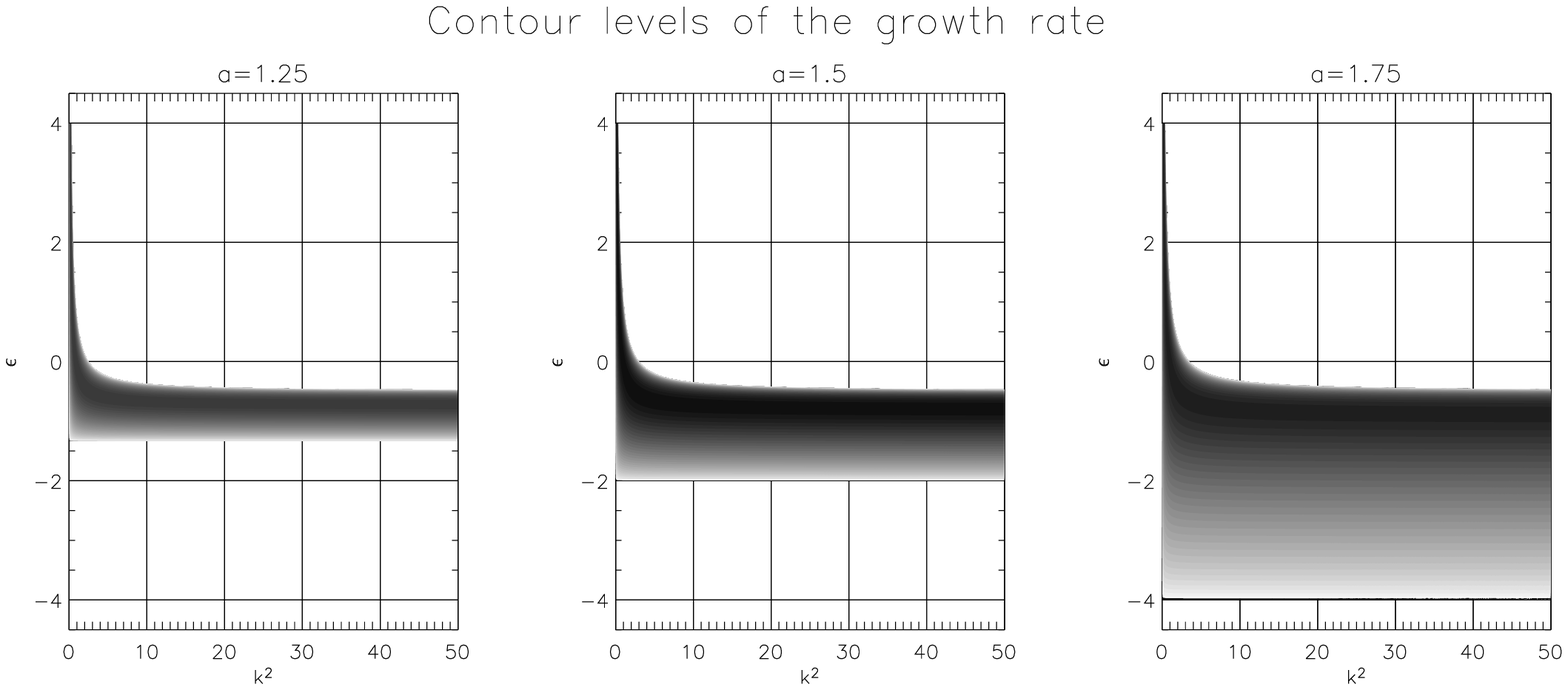}
\caption{Contour levels of the instability growth rate $\gamma_{-}$ on the
$(k^2, \varepsilon)$ plane for sub-keplerian ($s=1.25$), keplerian ($s=1.5$), and
super-keplerian ($s=1.75$) cases. Stable regions are white and darker shades correspond to larger instability growth rates.} 
\label{contours_hmri}
\end{figure}
%

%
Although we did not explore it in this paper, this instability also extends
toward small wavenumbers (i.e.\ $k \ll 1$ in our units). The maximum growth
rate $\gamma_{-}^{\max} = 0.5$ is actually attained for all parameter values
$\varepsilon$ and $k$ satisfying $(\varepsilon -2) k^2 = 1/2$. This particular
regime might correspond to the one considered by \citet{rudkit} for very large values of $\varepsilon$, which they termed as
``shear-Hall''. However, it must be pointed out that their study was
performed for differentially rotating plasmas. In the asymptotic limit of
very large $\varepsilon$ while $\varepsilon k^2$ remains constant, the effect of
rotation becomes gradually unimportant and the instability is driven by
the velocity
shear associated to the differentially rotating flow. To the best of our
knowledge, the Hall magneto-shear instability (as a process driven by a
mechanism unrelated with a differential rotation profile) was only
explored in its linear regime by \citet{kunz} for weakly ionized plasmas in a three-dimensional geometry.
\section{Hall Magneto-Rotational Instabilty (Hall-MRI)}\label{sec:hmri}
For the large-scale behavior of accretion disks many results have been obtained within the framework of magnetohydrodynamics (MHD). If the accretion disk displays differential rotation in the presence of magnetic fields, the magneto-rotational instability (MRI) appears as the most promising candidate \citep{velikhov, chandra, balhaw}. The magneto-rotational instability has been broadly tested by three-dimensional numerical simulations which show that the strong turbulence generates and enhances angular momentum transport efficiently \citep{bra, haw, mattaj}. However, kinetic plasma phenomena such as the Hall effect, might play an important role in the development of small-scale instabilities and turbulence. In particular, the influence of the Hall currents on the magneto-rotational instability has been considered in the linear \citep{wardle, balter, rudkit, devpek} as well as in the nonlinear regimes \citep{sanostone1, sanostone2}.

As shown in the previous section, in the presence of 
differential rotation the constant shear becomes $s = a\Omega = a\Omega_0 t_0$. In MRI studies it is customary to use $t_{0}=1/\Omega_{0}$, which in turn implies that $s=a$ (see for instance \citet{balhaw}). We adopt these units in the present section, so that our results can be compared with existing result in MRI in a straightforward manner. The characteristic length scale is
therefore $l_0 = v_A/\Omega_0$. In the presence of differential rotation, the coefficients of the 
dispersion relation are
\begin{equation}
C_{2}=\frac{\varepsilon ^{2}k^{4}}{2}+\left(1-\frac{\varepsilon a}{2}
\right)k^{2} +\left( 2 - a \right) \ ,
\label{eq:c2hmri}
\end{equation}
\begin{equation}
C_{0}= k^{2}\left[1+\varepsilon(2-a)\right]\left[k^{2}(1+2\varepsilon)-2a\right] \ .
\label{eq:c0hmri}
\end{equation}

Note that the cases where the rotation frequency $\mathbf{\Omega}$ is anti-aligned with the external magnetic field (i.e.\ $\Omega$ in Eqs.~(\ref{eq:c2})--(\ref{eq:c0})) can be described in Eqs.~(\ref{eq:c2hmri})--(\ref{eq:c0hmri}) with negative values of the Hall parameter $\varepsilon$. Since Eqs.~(\ref{eq:c2})--(\ref{eq:c0}) for $\Omega \neq 0$ and $s=a \Omega $, depend on the combination $\varepsilon \Omega $, with $\Omega = \pm 1$ (depending on the relative orientation of the rotation frequency with respect to the external magnetic field), we can also consider $\varepsilon \Omega $ as a ``signed $\varepsilon$''. In Hall-MRI, positive (negative) values of $\varepsilon$ correspond to $\Omega$ aligned (anti-aligned) to the external magnetic field.

We study three different rotation profiles: keplerian ($s = 1.5$), 
slightly sub-keplerian ($s = 1.25$), and slightly super-keplerian ($s = 1.75$). The growth rate $\gamma_-(k^2, \varepsilon)$ is shown in Fig.~\ref{growth_hmri},
for the keplerian case (i.e.\ $a=3/2$). Contour levels of this function are also
overlaid at the top as well as a gray scale image is shown at the bottom.

In Fig.~\ref{contours_hmri} we present the corresponding contour plots for the
following three cases: sub-keplerian, keplerian, and super-keplerian. Note that in all of
them, $\gamma_-$ reaches an asymptotic value which is positive and independent of $k^2$. Also, the whole unstable region becomes appreciably larger  when
differential rotation goes from sub-keplerian to super-keplerian regimes. An interesting feature of the Hall effect is the fact that it breaks the MHD symmetry ${\bf B}\leftrightarrow -{\bf B}$. A direct consequence of this symmetry breaking is a strong change of the instability region in the space of parameters depending on the relative alignment of the external magnetic field and the angular velocity of the disk. Also, the Hall effect qualitatively changes the criterion leading to instability even in cases where the angular velocity increases outward \citep{balter, sanostone1, sanostone2}.

The
instability region can be approximately described through a range of negative
values of $\varepsilon$, therefore corresponding to disks where the angular
velocity and magnetic field vectors are anti-aligned. For the cases where these
two vectors are aligned (i.e.\ $\varepsilon > 0$) instability only arises in a narrow region at the very
smallest wavenumbers allowed in the shearing box. The instability region in 
the $(\varepsilon , k^2)$ diagram, is confined within two curves of marginal stability, 
which are determined by the condition $C_0 = 0$, i.e.\
\begin{equation}
\varepsilon = - \frac{1}{2} + \frac{a}{k^2} \ ,
\label{eq:marg1}
\end{equation}
\begin{equation}
\varepsilon = - \frac{1}{2-a} \ .
\label{eq:marg2}
\end{equation}
For instance, for keplerian rotation  (i.e.\ $a = 3/2$) the instability strip at large 
wavenumbers ranges from $\varepsilon = - 1/2$ (i.e.\ Eq.~(\ref{eq:marg1}) for $k\rightarrow\infty$) 
to $\varepsilon = - 2$ (see. Eq.~(\ref{eq:marg2})). As we go to super-keplerian regimes, Eq.~(\ref{eq:marg2}) 
indicates that the instability strip widens up considerably, especially when the rotation parameter $a \rightarrow 2$.

The particular case of the standard magneto-rotational instability corresponds to $\varepsilon = 0$, for which 
we recover the classical result \citep{balhaw}. The highest growth rate for the unstable branch at $k^2 = 15/16$ is $\gamma_{-}^{\max} = 0.75$ and the range of unstable wavenumbers
is restricted to $0 < k^2 < 3$. For the sub-keplerian and super-keplerian
cases, the highest growth rates at the same $k^2$ are respectively $0.625$ and $0.875$, although each maximum is achieved for different values of the Hall parameter. Also the maximum growth rate can be obtained from the calculation of the local Oort A value of the rotation profile (see for instance \citet{balter}).

As mentioned at the beginning of this section, the influence of the Hall effect in the context of accretion disks has been previously analyzed. For instance, we obtain the same results for $k\rightarrow\infty$ reported by \citet{wardle}, \citet{balter}, and \citet{devpek}. In particular, Figs.~\ref{growth_hmri} and \ref{contours_hmri} can be straightforwardly compared with figures 1 and 5 of \citet{balter}. Also the dispersion relation presented by \citet{rudkit} in their equation 7, in the ideal limit, corresponds to the coefficients shown on Eqs.~(\ref{eq:c2hmri})--(\ref{eq:c0hmri}).
\begin{figure}
\epsscale{.4}
\plotone{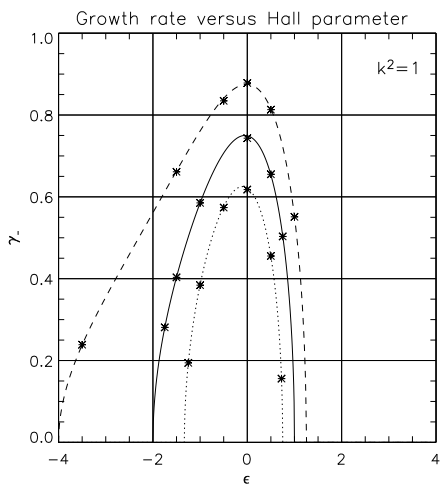}
\caption{Overlap between the theoretical model and the simulations
($\ast$) for $k=1$ and for three differential rotation profiles: sub-keplerian (dotted
line), keplerian (solid line), super-keplerian (dashed line). The well known
magneto-rotational instability corresponds to $\varepsilon=0$ where the highest
growth rate for the unstable branch is achieved.}
\label{overlap_hmri}
\end{figure}
\begin{figure}
\epsscale{.8}
\plotone{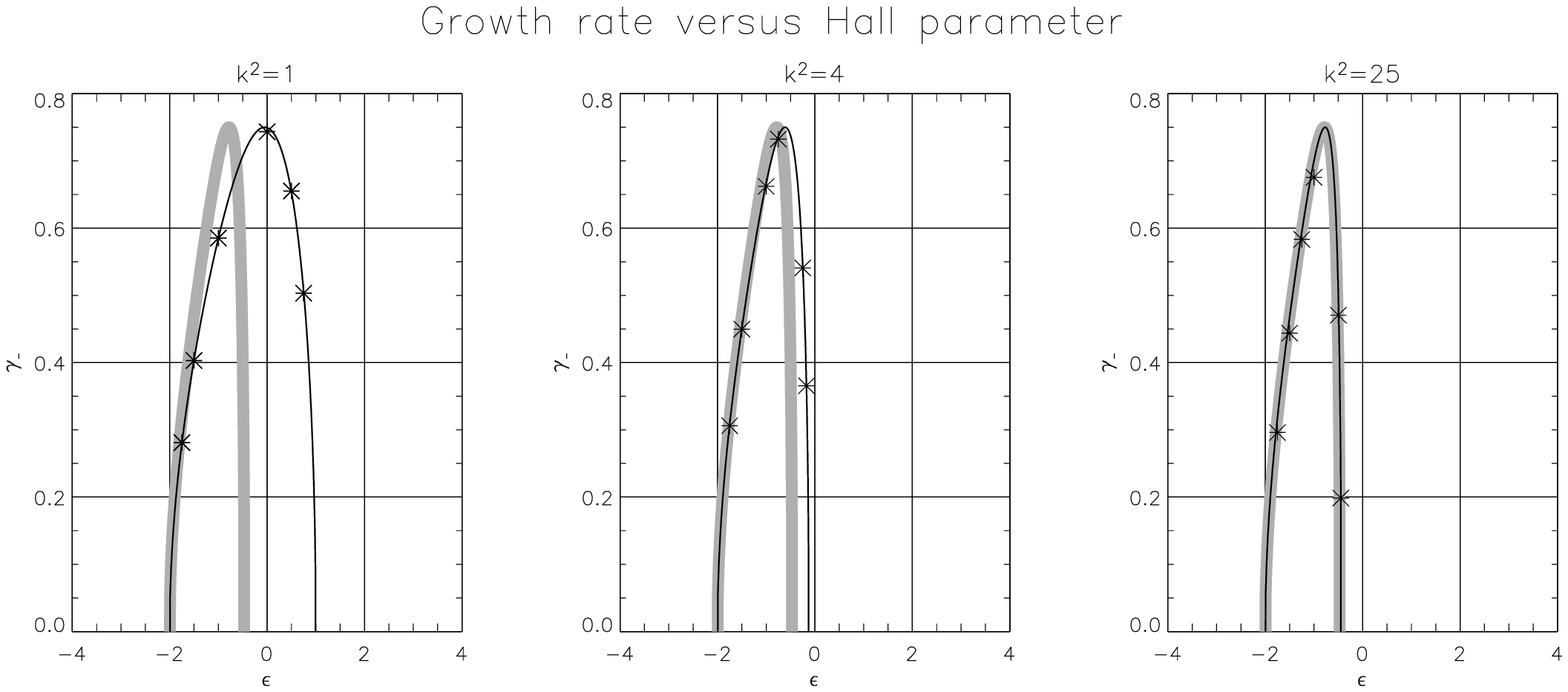}
\caption{Overlap between the theoretical model (solid line) and the
simulations ($\ast$) for $k=1$, $k=2$, and $k=5$.  The asymptotic behavior for
$k^{2} \rightarrow \infty$ (gray line) is indicated  in the three cases.}
\label{asym_hmri}
\end{figure}
%

%
In Fig.~\ref{overlap_hmri} we display transverse cuts of $\gamma_-$ vs. $\varepsilon$ for $k^2 = 1$. 
The three rotation profiles (i.e.\ sub-keplerian (dotted line), keplerian (solid line), super-keplerian 
(dashed line)) are overlaid for comparison. As we go to super-keplerian profiles, the instability 
range of values of $\varepsilon$ is broadened and the growth rates become progressively larger.
We performed several numerical simulations for different values of $\varepsilon$ and $a$, and the 
corresponding results are indicated with asterisks ($\ast$), showing a very good agreement with the 
analytical results. 
 
In Fig.~\ref{asym_hmri} we show three different cuts of $\gamma_-$ vs. $\varepsilon$ corresponding to three
different values of wavenumbers for the keplerian flow. Fig.~\ref{asym_hmri} exhibits a good correspondence between the numerical simulations and the analytical results. Note that the asymptotic
behavior (i.e.\ for $k\rightarrow\infty$) indicated with gray trace is gradually approched as the 
wavenumber increases. Just as for the Hall-MSI case in \S\ref{sec:hmsi}, the frequency of the unstable mode becomes gradually independent of $k$. In fact, the approximate balance between the convective and Hall terms on Eq.~(\ref{eq:linind}) is such as the large-$k$ asymptotic regime becomes independent of $k$.

The sub/super-keplerian profiles can be relevant in different astrophysical contexts. There are phenomenological constraints imposed on compact object (in particular, neutron stars and black holes) properties such as masses, spins, and size. The emitting regions of accretion disks are constrained by the assumption that super-keplerian oscillation frequencies cannot be observed from any radius. Therefore, the frequency of the lowest order linear hydrodynamic modes has to be smaller than the local keplerian value. The validity of this assumption is investigated by \citet{mao}. In \citet{kato09} the evolution of the magneto-rotational instablity is studied for weakly ionized protoplanetary disks, assuming a radially non-uniform magnetic field. For this physical configuration, the authors find that a zone with a super-keplerian velocity emerges as a result of the non-uniformly growing MRI turbulence. Likewise, in \citet{kato10}, the dust particles pile up at the boundaries of sub/super-keplerian regions,  in such a way that the dust density becomes large enough for the subsequent gravitational instability to set in. This result suggests a possible route to planetesimal formation from the dust particle in a protoplanetary disk  via the non-uniformly excited MRI. In accretion processes for which infalling matter does not really fall onto the disk surface, but onto the disk outer edge (which means that the disk accretes from the side instead of from the top), it may be necessary to consider deviations from keplerian rotation profile. In this context, the problem of sub-keplerian accretion disks requires particular attention \citep{cha, huegui, visdul}. A strong magnetization can make the disk surrounding young stellar objects to rotate at sub-keplerian rate (see for instance, \citet{shu} and \citet{paa}). 
\section{Non-linear behavior}\label{sec:nonlinear}
In the nonlinear regime, the Hall-MHD equations with the geometric configuration depicted by equation (\ref{eq:1d}) are as follows:
\begin{equation}
\partial_{t} \delta n  = -\partial_{z} u_{z} -\partial_{z}(u_{z} \delta n)
\ ,
\label{eq:nonlincont}
\end{equation}
\begin{equation}
(\partial_{t} +u_{z}\partial_{z})u_{z}  = -2 \beta \partial_{z} \delta n\ - \frac{1}{1+ \delta n} \partial_{z} \left( \frac{b_{\perp}^{2}}{2} \right) 
\ ,
\label{eq:nonlinnsz}
\end{equation}
\begin{equation}
(\partial_{t}+u_{z}\partial_{z})u_{\perp}  = \frac{1}{1+\delta n}(1+b_{z})\partial_{z}b_{\perp} + 
\left(
\begin{array}{cc}
0 & 2\Omega \\
(s-2\Omega) & 0
\end{array} 
 \right) 
u_{\perp}
\ ,
\label{eq:nonlinnsperp}
\end{equation}
\begin{equation}
(\partial_{t}+u_{z}\partial_{z})b_{\perp}  = -b_{\perp} \partial_{z}u_{z} +\partial_{z}u_{\perp}+
\left(
\begin{array}{cc}
0 & 0 \\
-s & 0
\end{array} 
 \right) 
b_{\perp}
+\varepsilon 
\left(
\begin{array}{cc}
0 & 1 \\
-1 & 0
\end{array}
\right) 
\partial_{z}\left((1+b_{z})\frac{\partial_{z}b_{\perp}}{1+\delta n}\right)
\ .
\label{eq:nonlinind}
\end{equation}

These non-linear equations show the interplay between the longitudinal (acoustic waves) and perpendicular modes (Hall-MSI or Hall-MRI). Even if the acoustic modes are initially turned off, the non-linear modes coupling with the perpendicular modes will eventually turn them on.

We analyze two different regimes according to the value of the
plasma parameter: $\beta \cong 1$ and $\beta \gg 1$ (high temperature). In the astrophysical context of accretion disks, the MRI instability  requires that the unstable wavelengths are smaller than the disk thickness, then the large $\beta$ regimes are therefore more relevant. However, we numerically verify that the non-linear regime for the very large beta is associated to the MHD configuration (i.e.\ without the Hall parameter).  

In \S\ref{sec:disprel}, the dispersion relation determined by Eqs.~(\ref{eq:c2})--(\ref{eq:c0}) shows that the linear behavior is independent of the plasma parameter, therefore
the instability growth rate is the same for both regimes because it is only a function of the wavenumber and the Hall parameter. The  root mean square ($f^{\rm rms}=\langle \mathbf{f}^{2} \rangle ^{1/2}$) of the proton velocity and the magnetic field  present very similar behavior for all cases, with different values of $\varepsilon$ and $\beta$ parameters. It should be noticed that,  if the growth rate increases,  the  duration of the linear period decreases, as expected. 

From equantions (\ref{eq:linnsperp})-(\ref{eq:linind}), the eigenvectors for the linear regime of the two instabilities (i.e.\ Hall-MSI and Hall-MRI) can be calculated. For Hall-MSI we obtain
\begin{equation}
\vel_{\perp}=\left(1, \frac{\varepsilon-(1+\gamma_{-}^{2})}{\gamma_{-} \varepsilon}\right)u_{x} \ ,
\end{equation} 
\begin{equation}
\cur_{\perp}=\left(\frac{(1+\gamma_{-}^{2})}{\varepsilon}, \gamma_{-} \right)u_{x} \ .
\end{equation} 
Also, it is possible to verify that
\begin{equation}
\tan{\phi_{u}}=\frac{\varepsilon-(1+\gamma_{-}^{2})}{\gamma_{-} \varepsilon}=\tan{\phi_{J}}=\frac{\gamma_{-} \varepsilon}{1+\gamma_{-}^{2}} \ .
\label{theta_hmsi}
\end{equation} 
Therefore, the proton velocity field and the current density are either aligned if the Hall parameter is positive or anti-aligned if the Hall number is negative (keeping in mind that the unstable modes are associated to positive values of the Hall parameter). 
For Hall-MRI the eigenvectors are
\begin{equation}
\vel_{\perp}=\left(1, \frac{\gamma_{-}(\gamma_{-}^{2}+1+\varepsilon(\varepsilon-a))}{2\gamma_{-}^{2}+2\varepsilon(\varepsilon-a)+\varepsilon} \right)u_{x} \ , 
\end{equation}
\begin{equation}
\cur_{\perp}=\left(\frac{\gamma_{-}^{2}(\gamma_{-}^{2}+1+\varepsilon(\varepsilon-a))}{2\gamma_{-}^{2}+2\varepsilon(\varepsilon-a)+\varepsilon}+\varepsilon-a,\frac{\gamma_{-}^{2}(\gamma_{-}^{2}+1+\varepsilon(\varepsilon-a))}{2\gamma_{-}^{2}+2\varepsilon(\varepsilon-a)+\varepsilon}-\gamma_{-} \right)\frac{u_{x}}{\gamma_{-}^{2}+\varepsilon(\varepsilon-a)} \ .
\end{equation}
For all values of the Hall parameter and any differential rotation profile (represented by $a$), we numerically show that the proton velocity field and the current density are always anti-aligned since $\cos(\theta_{u,j})=(\vel \cdot \cur) / |\vel||\cur| = -1$.
\begin{figure}
\epsscale{.6}
\plotone{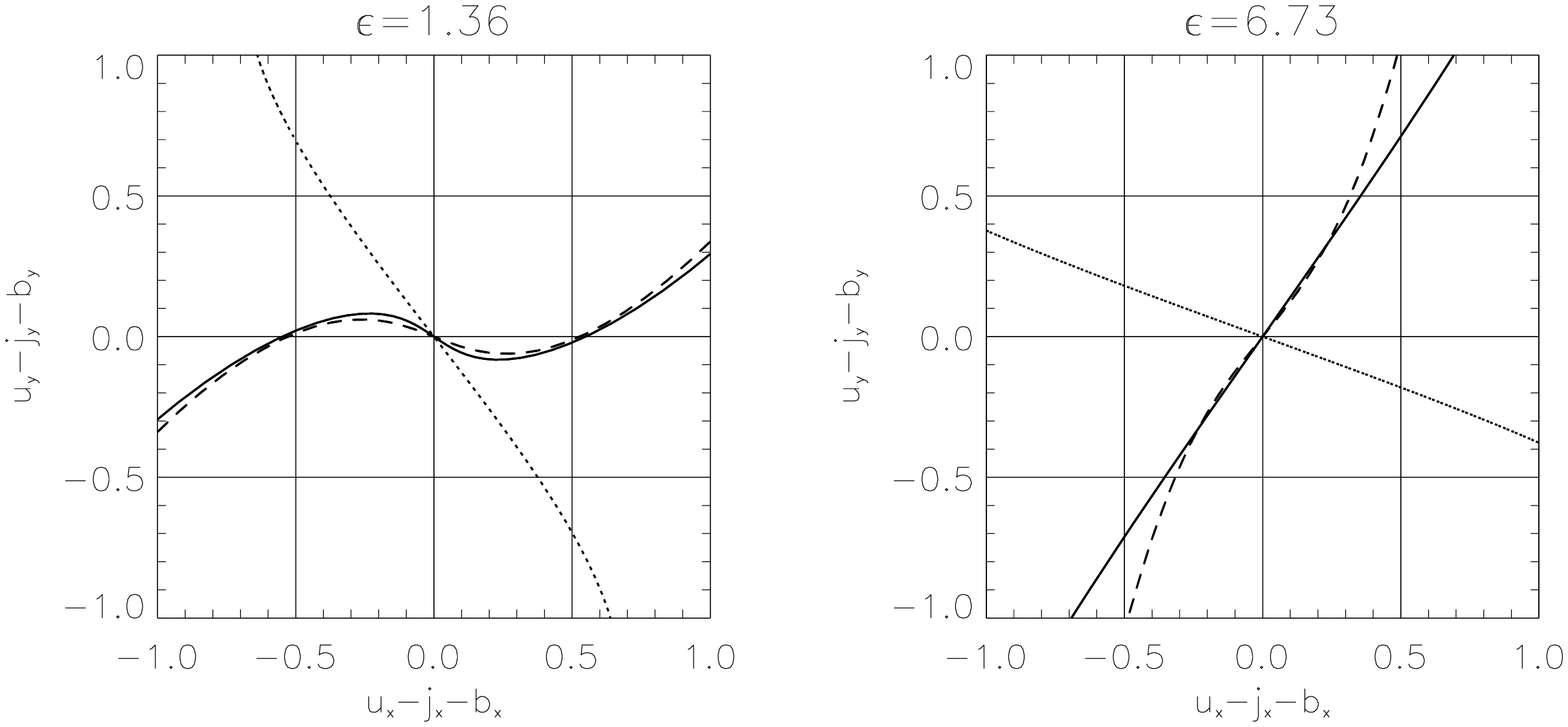}
\caption{Parametric representation of proton velocity (solid line), current density (dashed line), and magnetic field (dotted line) for Hall-MSI and for different values of the Hall parameter (labelled).}
\label{scatter_hmsi}
\end{figure}
\begin{figure}
\epsscale{.8}
\plotone{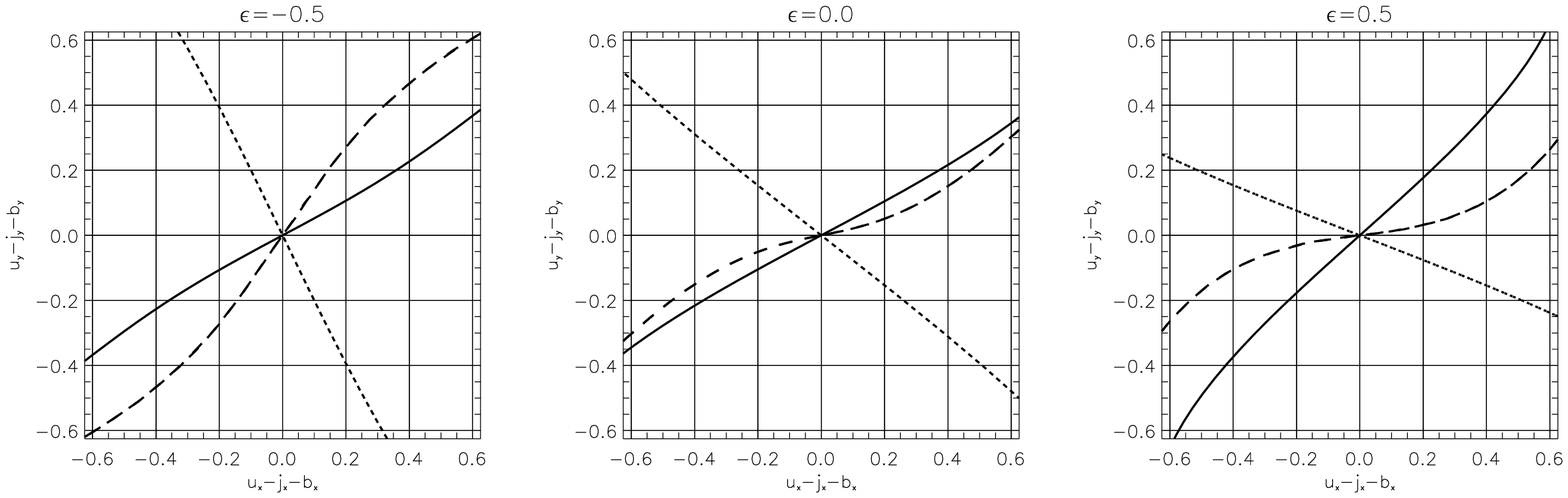}
\caption{Parametric representation of proton velocity (solid line), current density (dashed line), and magnetic field (dotted line) for Hall-MRI and for different Hall regimes: negative ($\varepsilon=-0.5$), MHD ($\varepsilon=0$), and positive ($\varepsilon=0.5$).}
\label{scatter_hmri}
\end{figure}
%

%
In order to give a qualitative description of the non-linear regime we select particular values of the Hall parameter. For Hall-MSI, we choose them in such a way that the growth rates correspond to $75 \%$ of the maximum, therefore we obtain $\varepsilon=1.36$ and $\varepsilon=6.73$. For Hall-MRI, we adopt values of the Hall parameter such that the corresponding growth rates are comparable to $90 \%$ of the maximum which takes place at $\varepsilon=0$ then we choose $\varepsilon=-0.5$ and $\varepsilon=0.5$. 

In Figs.~\ref{scatter_hmsi} and \ref{scatter_hmri} we display parametric plots of proton velocity (solid line), current density (dashed line), and magnetic field (dotted line) at times corresponding to the non-linear regime and for different values of the Hall parameter. From Fig.~\ref{scatter_hmsi}, which corresponds to Hall-MSI, it seems apparent that, regardless of the values of the Hall parameter, the eigenvectors for the velocity field and the current density are always aligned thus extending the result shown in Eq.~(\ref{theta_hmsi}) to the nonlinear regime as well. Similarly, Fig.~\ref{scatter_hmri} shows that for Hall-MRI the velocity field and the current density are always anti-aligned as we numerically find in the linear regime. 
 
In both cases, we note that the influence of the non-linear terms becomes non-negligible whenever the density fluctuations reach about ten percent of the mean value. In particular, the regime becomes strongly nonlinear when the velocity is equal to the Alfv\'en speed. In the super-alfv\'enic regime, for Hall-MSI with strong Hall effect the whole system seems to start reaching a saturation state and for Hall-MRI with negative of the Hall parameter the system  seems to reach saturation.
 
Also, the field configurations in the non-linear stage depend on the value of the Hall parameter. In the Hall-MRI case for $\varepsilon < 0$, the $x$-component of the current density is larger than the $y$-component, while for $\varepsilon > 0$ the inverse situation occurs. In both cases, the system shows the formation of the current sheet in the larger component. For the MHD case, of course, the electron and proton velocities are the same. In this case, current sheets are present in the $\mathbf{\hat{x}}$ and $\mathbf{\hat{y}}$ directions. In the Hall-MSI case with moderate Hall parameter value, current sheets seem to appear in both directions.

\begin{figure}
\epsscale{.6}
\plotone{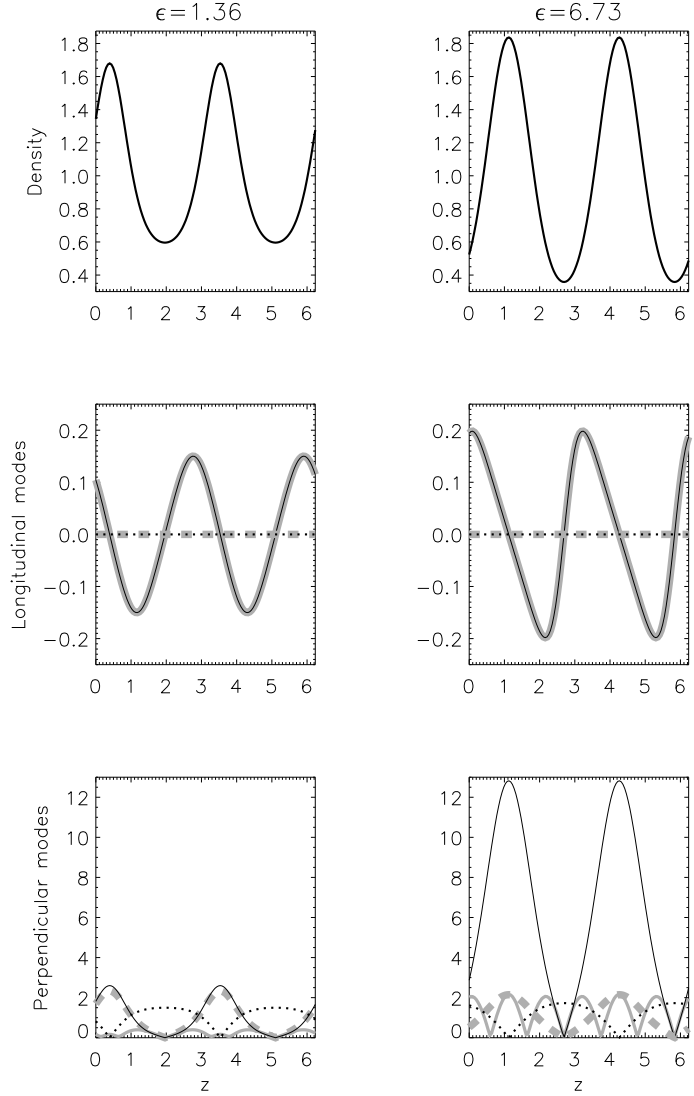}
\caption{Non-linear behavior of Hall-MSI for $\beta=1$ and for two different values of the Hall parameter (labelled). The upper panels show the particle density, the central panels display the longitudinal modes (i.e.\ the $z-$components), and the lower panels show the perpendicular modes (i.e.\ $(x,y)$-components). There are four vector fields represented: proton (solid black line) and electron (solid gray line) velocities, current density (dashed gray line), and the magnetic fields (dotted black line).}
\label{profiles_hmsi}
\end{figure}

\begin{figure}
\epsscale{.8}
\plotone{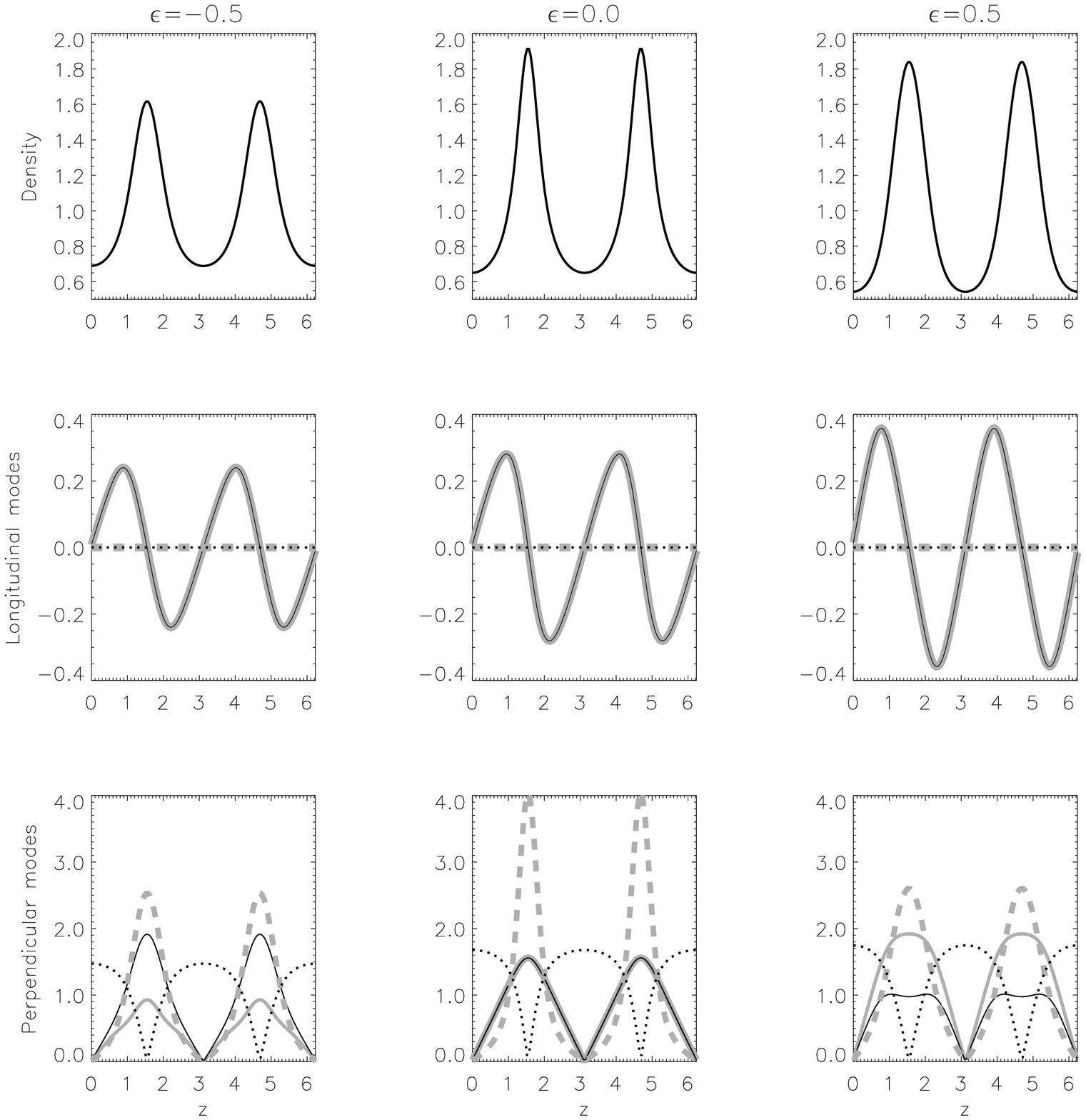}
\caption{Non-linear behavior of Hall-MRI for $\beta=1$ in three different Hall regimes: negative ($\varepsilon=-0.5$, left panels), MHD ($\varepsilon = 0$, central panels), and positive ($\varepsilon=0.5$, right panels). In each regime, the figure displays the density (upper panels), the longitudinal (central panels) and the perpendicular modes (lower panels).  There are four vector fields represented: proton (solid black line) and electron (solid gray line) velocities, current density (dashed gray line), and the magnetic fields (dotted black line). }
\label{profiles_hmri}
\end{figure}

In Figs.~\ref{profiles_hmsi} and \ref{profiles_hmri} the profiles of the density and the vector fields $\vel$ (solid black line), $\vel_{e}$ (solid gray line), $\cur$ (dashed gray line), and $\magfp$ (dotted black line) are shown in the nonlinear regime as functions of $z$ for Hall-MSI and for Hall-MRI, respectively. Negative (positive) values of the Hall parameter imply that the electron velocity is always smaller (larger) than the proton velocity. A more extreme situation takes place in the Hall-MSI case being the proton velocity very much larger than the electron velocity. In particular, for strong Hall effect this is clearly shown in Fig.~\ref{profiles_hmsi} (lower panels). In addition, the Hall-MRI case for positive values of the Hall parameter seems to show the formation of two-flow jets. These double jet structures have also been reported by \citet{sanostone1} in 2D simulations. It is relevant to note that the $z$-component of the velocities start to increase when the density fluctuations grow and the non-linear regime is reached. However, this component of the velocities (for protons and electrons) remains negligible. Figs.~\ref{profiles_hmsi} and \ref{profiles_hmri} also show that the longitudinal modes for protons and electrons are exactly overlapped. Meanwhile, the $z$-component of the magnetic field and the current density are approximately zero.
%

%
Moreover, the maximum values of the density are correlated with the maximum of the perpendicular velocities and with the maximum of the current density, i.e.\ zero magnetic field. Meanwhile, the minimum values of density are in relation with the zeroes of the velocity fields and current density, i.e.\ maximum magnetic field (see Figs.~\ref{profiles_hmsi} and \ref{profiles_hmri}). 

Even though the nonlinear regimes discussed in this section show interesting features, many of them might change when we extend our simulations to three dimensions. The purpose of this 1D study was to identify different regimes in the space of parameters of the problem.
\section{Conclusions}\label{sec:conclu}
The present work is a comprehensive study of the instabilities arising from the interplay between the Hall effect and a linear shear flow for a one-dimensional model. In other words, we analyze the role of the Hall effect in shear-driven instabilities. We find that an instability develops when the Hall effect is present, which we term Hall magneto-shear instability. Also, we recover the magneto-rotational instability, as a particular case, and we quantitatively evaluate the influence of the Hall currents on it.

More specifically, we investigate the stability of the system in the parameter space set by the wavenumber and the Hall parameter. In non-rotating plasmas, we determine the region in the $(k^{2}, \varepsilon)$ diagram where the Hall magneto-shear instability takes place. In rotating plasmas (such as accretion disks), we examine three cases: sub-keplerian, keplerian, and super-keplerian. For each rotation profile, we establish the region in the $(k^{2}, \varepsilon)$ diagram where the Hall magneto-rotational instability occurs. The standard MRI is recovered in the particular case for zero Hall parameter ($\varepsilon=0$). In both unstable modes, we find a very good agreement between the theoretical model and the numerical simulations.

In addition, we explore the influence of the plasma parameter in two asymptotic cases: $\beta \cong 1$ and $\beta \gg 1$. We find that the linear behavior is independent of this parameter. In the non-linear stage of the large beta regimes, the flow dynamics seems to evolve like in an MHD system. 

Whitin the framework of astrophysics, the Hall magneto-shear instability could be relevant in the interface between a jet and the surrounding environment where a strong shear is present. Astrophysical jets have a very high degree of collimation, probably as a consequence of the presence of magnetic fields. The one-dimensional model adopted in this work, even though quite simple, might properly describe this circumstance: the azimuthal component of the helicoidal velocity field arround the jet can be represented as a function of the radial direction and fulfills the periodicity condition in a straightforward fashion. The relevant instability in these strongly sheared flows is Kelvin-Helmholtz, which is a purely hydrodynamic instability. The presence of external magnetic fields modify the corresponding growth rate, depending on their strength and spatial orientation, but typical numbers quoted in the literature remain a small fraction of the imposed velocity shear (\cite{ferrari, bodo} and \citep{huba} for results from Hall-MHD simulations). Therefore, the Hall magneto-shear instability, with maximum groth rate of $0.5$ times the externally imposed shear, is definitely relevant in these strongly sheared flows. On the other hand, it seems clear that a three dimensional extension of the present study is necessary for a more realistic description of this instability, especially when it comes to its nonlinear stage.

\acknowledgments

This work has been supported by the University of Buenos Aires through grant
UBACyT X092/2008 and by the ANPCyT through grant PICT 33370/2005.



%
%

\end{document}